\documentclass[prb,aps,floatfix,amsmath,amssymb,superscriptaddress]{revtex4}

\usepackage{mathtools}
\usepackage{amsfonts}
\usepackage{amssymb}

\usepackage{graphicx}
\usepackage[all]{xy}

\usepackage{amsthm}

\newcommand{\dt}{{\rm d}t}
\newcommand{\ini}{\rm initial}
\newcommand{\fin}{\rm final}

\newcommand{\be}{\begin{equation}}
\newcommand{\ee}{\end{equation}}

\begin{document}
\title{Connecting Entanglement in Time and Space: Improving the Folding Algorithm}
\author{M. B. Hastings}
\affiliation{Microsoft Research, Station Q, CNSI Building, University of California, Santa Barbara, CA, 93106}
\affiliation{Quantum Architectures and Computation Group, Microsoft Research, Redmond, WA 98052, USA}

\author{R. Mahajan}
\affiliation{Department of Physics, Stanford University, Stanford, CA, 94305, USA}

\begin{abstract}
The ``folding algorithm"\cite{fold1} is a matrix product state algorithm for simulating quantum systems that involves a {\it spatial} evolution of a matrix
product state.  Hence, the computational effort of this algorithm is controlled by the {\it temporal} entanglement.  We show that this temporal entanglement is,
in many cases, equal to the spatial entanglement of a modified Hamiltonian.  This inspires a modification to the folding algorithm, that we call the ``hybrid
algorithm".  We find that this leads to improved accuracy for the same numerical effort.  We then use these algorithms to study relaxation in a transverse plus parallel field Ising model, finding persistent quasi-periodic oscillations for certain choices of initial conditions.
\end{abstract}
\maketitle

In many cases, one-dimensional quantum systems can be accurately simulated on a classical computer.  For ground state properties, the density-matrix
renormalization group algorithm (DMRG)\cite{dmrg} is often very accurate in practice.  This algorithm relies crucially on the ability to approximately represent quantum
states in one-dimension with low entanglement entropy in terms of matrix product states (MPS), and for many physical systems this low
entanglement property holds for the ground state.

For time dynamics, the problem becomes significantly more complicated.  The time-evolving block decimation (TEBD)\cite{tebd} and related time-dependent DMRG\cite{tdmrg} work well so long as the entanglement of the state remains small during time evolution. This entanglement indeed grows slowly enough for many physically interesting problems to obtain useful results (see references and the review in Ref.~\onlinecite{dmrgmps}).  However, for other interesting problems involving a global quench, the entanglement grows linearly in time\cite{entropy}, leading to
an exponential growth of the bond dimension required to simulate these systems using such matrix product methods.  This growth is within a constant factor
of the maximum possible\cite{eb1,entropybound,eb2,eb3}.

A large entanglement entropy indeed should be expected for systems that thermalize after a quench. Thermalization after a quench refers to a situation in
which a closed quantum system is subject to a sudden global change in the Hamiltonian such that the local reduced density matrices converge to thermal
density matrices.  Since the thermal density matrices have entanglement entropy that is linear in the size of the region at nonzero temperature, this leads to an entanglement entropy that is at least proportional to the length scale over which the system has thermalized.
This property is somewhat paradoxical: even if the system is approaching some state that is relatively simple with only short range correlations (a thermal state at nonzero temperature in one-dimension), the description using matrix product methods is extremely complicated because while the local state does become mixed the global state remains pure.

However, one potential route around this is to evolve the system {\it transversally}, i.e. in space rather than in time, as suggested in Ref.~\onlinecite{fold1}.  
This method involves using a matrix product state to describe the effective time dynamics of a single spin, after tracing out the spins to the left (or right) of that given spin.  The efficiency of this method depends then on the ``entanglement in time" of that state.
In this paper, we consider the entanglement in time in a continuum limit and show that it is very closely related to the spatial entanglement of a modified Hamiltonian.  For real-time dynamics, this Hamiltonian is non-Hermitian.  We analyze the entanglement of this non-Hermitian Hamiltonian in some simplified settings and analyze the entanglement growth.  We then propose a natural modification of the folding method, that we term the ``hybrid method", inspired by this study of the entanglement.  We study the performance of this hybrid method numerically; in every situation we analyzed, it reduces the error compared to the original folding method.  We then use the method to study a one-dimensional spin chain with slow relaxation; this chain is the same as that
considered in Ref.~\onlinecite{fold2}, but we consider other possible initial conditions and show in this case that long-time oscillations persist even in single-body observables for times as far as we can analyze numerically.  The failure to thermalize for a long range of times is even stronger than the
results in Ref.~\onlinecite{fold2}, presenting further evidence that the isolated quantum systems may {\it not} thermalize in general, even if many cases do\cite{therm3,therm2,therm1,therma,therm0,thermf}.
The hybrid method can be understood without reading the earlier sections, though they provide some useful motivation and help fix notation.

Let us begin with an overview of the folding method and more generally of methods of transverse evolution.
Suppose we wish to evaluate the expectation value of some local observable $O$ at some time $T$, using an initial state $\Psi$ and Hamiltonian $H$, so that we wish to compute $\langle \Psi | \exp(i H T) O \exp(-i H T)| \Psi \rangle$.
Using a Trotter-Suzuki decomposition of $\exp(-i H \dt)$ for small $\dt$, it is possible to approximate this expectation value as a two dimensional tensor network.  Each row of the tensor network is a matrix product operator (MPO) that represents the evolution forward in time by one time step.  The usual TEBD algorithm works roughly as follows: imagine cutting the tensor network along some horizontal line which cuts through a row of vertical bonds.  The tensor network below this line describes some state on these bonds; approximate this state as an MPS of given bond dimension $\chi$.  Then, applying the MPO representing a single timestep to this MPS gives a new MPS describing the state with the horizontal line moved up by one column.  The bond dimension of this new MPS is larger; if it
exceeds some given maximum value, this MPS is then truncated to a new MPS with smaller bond dimension.  Iterating in this way, one obtained a matrix
product description of the state as a function of time.  One could apply this algorithm to the entire network (moving forward by time $t$, applying $O$, then moving backward by time $t$), but the usual TEBD algorithm exploits the fact that the initial and final states are identical to simply evolve the state at the bottom of the network forward by time $t$ (i.e., through half the tensor network), then evolve the state at the top of the network backward by time $t$ (in fact, this evolution does not need to be done explicitly since this state is simply the conjugate of the state at the bottom), and then compute the expectation value of $O$ sandwiched between these two MPS.

The method of Ref.~\onlinecite{fold1} instead evolves sideways.  Naively, one would do precisely the same thing as described above, except instead of cutting the network on some horizontal line and evolving upward, one cuts the network on some vertical line and evolves to the right.  If $O$ is localized on some small set of sites, one can evolve one state from the left and one form the right and compute the expectation value of $O$ this way.
The performance of this method then will depend not on the entanglement in {\it space} but rather on the entanglement in {\it time}.
We may hope that this method will then perform well in some situations where TEBD fails.  In particular, for problems of thermalization where the
spatial entanglement is quite large, this method offers an alternative.

In fact, this method as described here does not offer much, if any, advantage.  However, as described in Ref.\onlinecite{fold1}, it is possible to improve this method if one {\it folds} the tensor network about a horizontal line drawn through the middle of the network.  This combines pairs of sites at the same time (one site  being at that time above the middle of the network where evolution occurs backwards and the other being below the line where the evolution occurs forward) into a single site of larger dimension.  Then, this folded state is evolved transversally.
In this way, the entanglement entropy is often reduced, allowing access in some cases to a description of physical systems at times beyond what can be accomplished with TEBD\cite{fold1,fold2,fold3}.  Physical intuition for the success of this method is offered in those references; another way to think about it is to imagine the evolution to the right.  Then, the spins to the left of a given column induce some dynamics on that column of spins.  If this dynamics is in some sense ``noisy", it will make that column of spins behave classically, with the spin being some classical variable with (for large noise) little correlation from one time instant to the next.  This spin would then be maximally entangled between pairs of spins that are joined by the fold, so the entanglement entropy would be large before folding but small after folding.

\begin{figure}
\includegraphics[width=2.5in,angle=270]{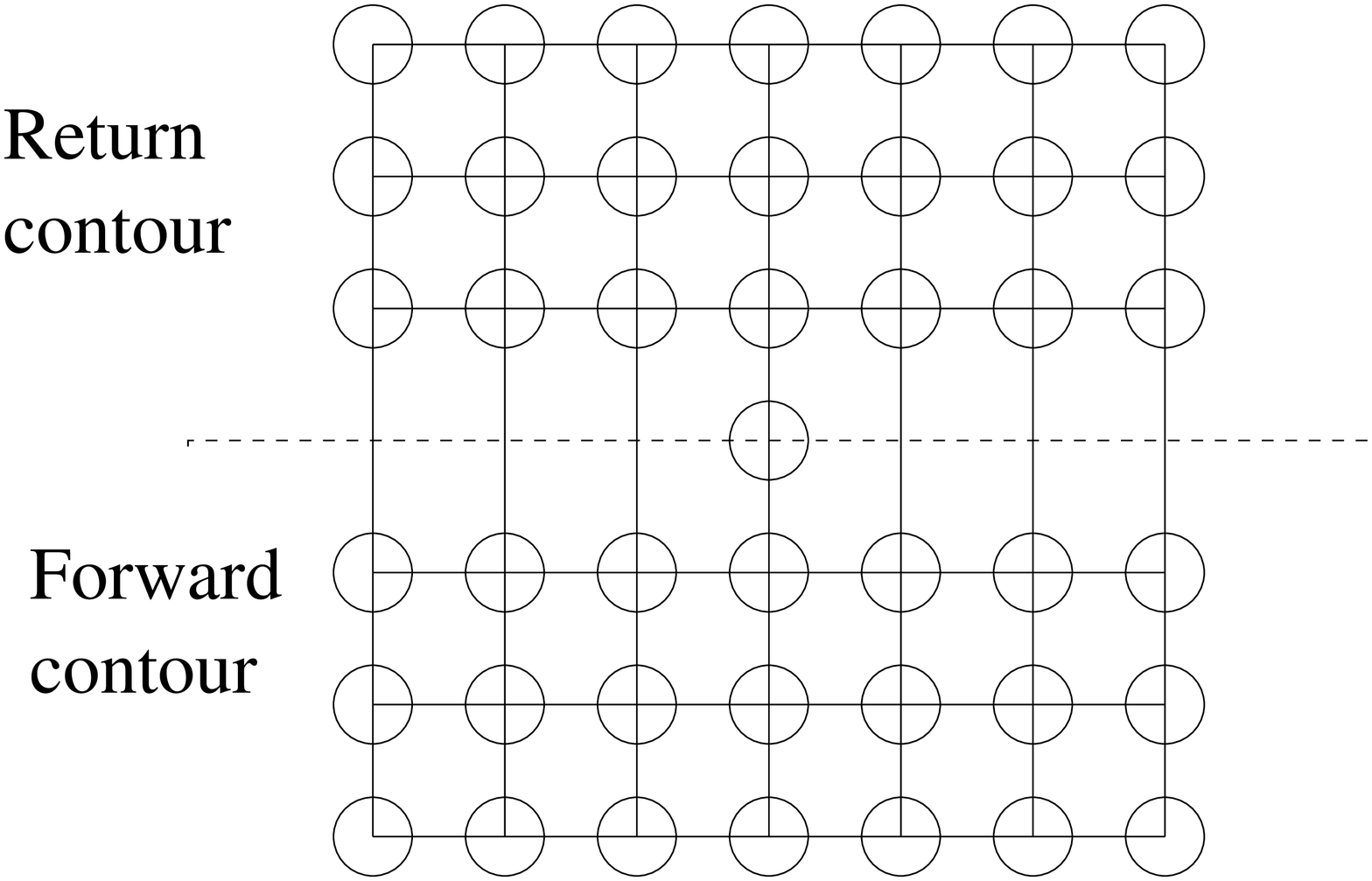}
\caption{Figure showing tensor network.  Observable is at middle of tensor network.  Bottom half is forward evolution in time, called ``forward contour",
while top half is backwards evolution in time, called ``return contour".  States at top and bottom are matrix product states.
We show two Trotter-Suzuki steps on each contour.  There are a total of $7$ sites.  The operator at the middle of the network represents
an observable on the fourth site.}
\label{fignet}
\end{figure}

We now describe the tensor network used in the discrete version of the problem to compute the expectation value
$\langle \Psi | \exp(i H T) O \exp(-i H T)| \Psi \rangle$ in more detail.  See Fig.~\ref{fignet}.  First consider the unfolded case.

At the bottom and top rows of the tensor network, we have the initial and final states $|\Psi\rangle$ and $\langle \Psi|$, described as matrix product states.
We choose some discrete time step $\dt$.  
For definiteness, consider the Hamiltonian
\be
\label{Ham1}
H=\sum_i JS^z_i S^z_{i+1} + g S^z_i + H S^x_i.
\ee
More general Hamiltonians are possible and we describe those below, however this will be the Hamiltonian studied numerically in this paper.

We will describe the evolution using a second-order Trotter-Suzuki decomposition, writing
$H=A+B$, 
where $A=\sum_i  J S^x_i+ g S^z_i$ and
$B=\sum_i JS^z_i S^z_{i+1}$,
approximating $\exp(-i H \dt) \approx \exp(-i A \dt/2) \exp(-i B \dt) \exp(-i A \dt/2)$.
Note that $\exp(-i A \dt/2)$ can be written as a matrix product operator of bond dimension $1$ (indeed, it is a product of unitaries on each site), while
$\exp(-i B \dt)$ can be written as a matrix product operator of bond dimension $2$ as in Ref.~\onlinecite{mporep}.
Hence, a single Trotter-Suzuki step can be written using a matrix product operator of bond dimension $2$.  This is the decomposition we use
for the rows of the tensor network, while at time $t$ we insert the operator $O$.  This operator breaks the translation invariance of the system in space.

More generally, we can decompose any interaction using these kinds of matrix product operators.  For example, if in addition we have an
 $XX$ interaction $C=\sum_i K S^x_i S^x_{i+1}$, we can also write $\exp(-i C \dt/2)$ as a matrix product operator of bond dimension two, so that
the evolution under $H=A+B+C$ can be written using matrix product operators of bond dimension $2 \times 2=4$.

We refer to the bottom of the network as the ``forward contour" and the top half of the network as the ``return contour".
For condensed matter physicists, the folding method can be viewed as a ``Keldysh" method, where these terms forward and return originate.
The folding operation combines a pair of sites, one on the forward contour and one on the return contour, into a single site.
After folding, the bond dimension of the matrix product operators becomes squared, so that for $H=A+B$ we need bond dimension $4$.

\section{Transverse Evolution in the Continuum Limit}
In this section we analyze transverse evolution in the continuum limit.  We consider variety of cases, included real-time and imaginary time as well as folded an unfolded.  The main result is to show that the entanglement can be considered by analyzing a modified Hamiltonian.

\subsection{Continuum Limit}
Under the transverse evolution above, we use a matrix product state to describe the state of the system on a column of horizontal bonds, after summing over the tensor network to the left (or to the right) of that column; we call such a state a left (or right) {\it transverse state}.  Consider first the unfolded case and the Hamiltonian Eq.~(\ref{Ham1}) for real time dynamics (the basic ideas in this subsection carry over with slight modifications to imaginary time dynamics as well as the folded case and we simply pick this case for definiteness).   We write the matrix product operator $\exp(-i B \dt)$ as
\be
\label{product}
\prod_{i} \exp(-i J S^z_i S^z_{i+1} \dt) = \prod_i \Bigl(\cos(J \dt) - i \sin(J \dt) S^z_i S^z_{i+1}\Bigr).
\ee
We write this as a matrix product operator where each bond can be in one of two states that we call $I$ or $Z$, corresponding to the
first and second terms in the product of Eq.~(\ref{product}).

Label each bond by a time $t$ and by a value $\sigma=f$ or $\sigma=r$, where $\sigma=f$ for the bonds on the bottom half of the network and $\sigma=r$ for those on the top half of the network.  This notation $f,r$ refers to ``forward" or ``return", using the notation above.  We specify a bond by a pair $(t,\sigma)$, and give an order on these bonds, ordering them from bottom to top, so that $(t,r)$ is after $(t',f)$ for all $t,t'$ while $(t,f)$ is after $(t',f)$ if $t>t'$ and $(t,r)$ is after $(t',r)$ if $t<t'$ (note the different direction of the inequality on the return contour).  We write $(t,\sigma)>(t',\sigma')$ if $(t,\sigma)$ is after $(t',\sigma')$.

After summing over spins to the left of a given column of bonds, we obtain some left transverse state.  
Rather than describing this state by giving the value of each bond, we instead describe it by giving the number of bonds in the $Z$ state and by listing the pairs $(t_1,\sigma_1), (t_2,\sigma_2), ..., (t_n,\sigma_n)$ for which a bond is in a $Z$ state, ordering so that $(t_1,\sigma_1)<(t_2,\sigma_2)<...$.
Additionally, if the initial and final states $\Psi$ are described by matrix product states, we specify the bond values $\alpha_{\ini}$ and $\alpha_{\fin}$ on the bonds on the bottom and top rows; we write $\Psi=\sum_\alpha \Psi_L^\alpha \Psi_R^\alpha$, where $\Psi_L,\Psi_R$ are states on the spins to the left and right of the given column, respectively.
Consider now the limit of small $\dt$.  Sum over spins $1,...,N$, fixing the cut between spin $N$ and spin $N+1$.  We can then exactly compute the amplitude to have a given sequence $(t_a,\sigma_a)$ and a given $\alpha_{\ini},\alpha_{\fin}$ on the bonds connecting spin $N$ to spin $N+1$.  It equals
\be
(J \dt)^{n/2} \langle \Psi_L^{\alpha_{\fin}}|  S^z_N(t_n) S^z_N(t_{n-1}) ... S^z_N(t_1) | \Psi_L^{\alpha_{\ini}} \rangle,
\ee
where $S^z_N(t)=\exp(it H_L) S^z_N \exp(-i t H_L)$ and $H_L$ is the Hamiltonian on spins $1,...,N$.
The factor of $(J\dt)^{n/2}$ is chosen so that the contraction of two matrix product states, one on the left and one on right of the given column,
gives a factor $(J\dt)^n$, correctly weighting the interactions.

The crucial point is that in the limit $\dt \rightarrow 0$, this amplitude describes a {\it continuous matrix product state}\cite{CMPS}.
This continuous matrix product state has the special property that $H_L$ is Hermitian.  However, we will see that gauge transformations (necessary to accurately truncate the problem) will lead to violations of this property.

Now consider the entanglement of this matrix product state across some cut at pair $(t,\sigma)$.  Note that if there are $N$ spins to the left of the given column, labelled $1,...,N$, then the matrix product state has bond dimension at most $2^N$ and the bond variables are simply the configurations of those $N$ spins.  Thus, we can label the bond variables by any orthonormal basis for those spins.
Suppose those spins are in some state $\alpha$.  This then corresponds to some state of the bond variables above and below the pair $(t,\sigma)$.  We need to compute the matrix $\Lambda_b$ whose matrix elements $(\Lambda_b)_{\alpha\beta}$ give the inner product of the state of bond variables for a given pair, $\alpha,\beta$ for the bonds below the given cut.  We then need to compute the matrix $\Lambda_t$ giving the inner product for the bond variables above the cut.
Having computed these two matrices, we can compute the entanglement as follows.
Consider the matrix
\be
\Lambda \equiv \Lambda_t^{1/2} \Lambda_b \Lambda_t^{1/2}.
\ee
Normalize this matrix $\Lambda$ by multiplying by a constant so that it has trace $1$.  It is a positive definite matrix and the entanglement entropy is simply the entropy
of that matrix $\Lambda$ regarded as a density matrix.

So, we now consider the matrix $\Lambda_b$.
We can write an equation of motion for $\Lambda_b$.  The initial conditions at $t=0, \sigma=f$ are determined by the inner product of the states
$\Psi_L^{\alpha}$.  The equation of motion is given by (for $\sigma=f$)
\be
\partial_t \Lambda_b=i [ \Lambda_b,H_L] +|J| S^z_N\Lambda_b S^z_N.
\ee
Note the absolute value sign around $J$.
For $\sigma=r$, the equation changes to $\partial_t \Lambda_b=i [ \Lambda_b,H_L] -|J| S^z_i \Lambda_b S^z_i$; note that for $\sigma=r$ the evolution upwards is in the direction of decreasing $t$.  The equations of motion for $\Lambda_t$ are the same except that the sign of the first term on the right-hand side is changed.

This equation can be regarded in one of two different ways.  We can regard $\Lambda_b$ as a density matrix.  Then, note that
\begin{eqnarray}
\label{timeev}
\partial_t \Lambda_b &=& i [ \Lambda_b,H_L] +|J| S^z_N \Lambda_b S^z_N \\ \nonumber &=& i [\Lambda_b,H_L] -|J| [[\Lambda_b,S^z_N],S^z_N]+2|J|\Lambda_b.
\end{eqnarray}
The last term on the right-hand side of the last line only changes the overall normalization of $\Lambda_b$ so it may be dropped when considering the entanglement.  Then, the first two terms describe the dynamics of a system under Hamiltonian evolution with $H_L$ and additional dephasing of the $N$-th spin.
For generic choices of $H_L$, the unique fixed point is the identity matrix.

Alternately, we can regard $\Lambda_b$ as a pure state on a system of size $2N$.  In this case, it describes a system with Hamiltonian $+H_L$ on the left half, $-H_L$ on the right-half, and a {\it non-Hermitian} coupling between the two halves.  For such a pure state $\Phi$ on a system of size $2N$
we write
\be
\label{Phieq}
\partial_t \Phi =\tilde H \Phi,
\ee
where
\be
\tilde H= -i (H_L \otimes I - I \otimes H_L^{ref}) \Phi + |J| S^z_N S^z_{N+1},
\ee
labelling the spins $1,...,2N$, and calling the first $N$ spins the left half and the second $N$ spins the right half.  The term $H_L \otimes I$ denotes the Hamiltonian $H_L$ acting on the left half, while the Hamiltonian $H_L^{ref}$
denotes the Hamiltonian $H_L$ reflected in space, i.e., $H_L^{ref}=R H_L R$, where $R$ interchanges the state of spins $k$ and $N+1-k$.
For the particular Hamiltonian we consider, we have $H_L=H_L^{ref}$ as it is reflection symmetric in space.

This general procedure can be carried out also for folded systems.  In this case, if we sum over the first $N$ spins, the matrix product state
has bond dimension at most $4^N$.  The bonds are now labelled by a single time $t$ as both contours are moved into one bonds.
Each bond can be in one of {\it three} states, that we label $I, Z_f,Z_r$, where $Z_f$ corresponds to an insertion of $S^z_N$ on the forward
contour while $Z_r$ corresponds to an insertion of $S^z_N$ on the return contour.
Note that this is one advantage of the continuum limit: the amplitude to insert $S^z_N$ on both forward and return contour at the same time vanishes
as $\dt \rightarrow 0$.  More generally, if the unfolded problem has bond dimension $d$, then for a discrete problem the bond dimension becomes $d^2$ while for the continuum limit the bond dimension is only $2d-1$.

\subsection{Imaginary Time}
The analysis is similar in imaginary time.  In this case, we have
\begin{eqnarray}
\partial_t \Lambda_b &=& - \{ \Lambda_b,H_L\} +|J| S^z_N \Lambda_b S^z_N.
\end{eqnarray}
Regarding $\Lambda_b$ as a pure state on a system of size $2N$, as in Eq.~(\ref{Phieq}), this describes the imaginary time evolution of the same Ising model as the original
Hamiltonian, with the exception of the fact that the coupling between spins $N$ and $N+1$ is by $|J|$ rather than $J$ and hence is always ferromagnetic.

For large times, then, the state $\Phi$ converges to the ground state of this Hamiltonian.  Further, in this case we have $\Lambda_b=\Lambda_t^\dagger$, so
the entanglement entropy of the state $\Lambda$ is {\it identical} to the spatial entanglement entropy of the state $\Phi$ between the two different halves.

Now consider an algorithm that describes the transverse evolution of this state.  Such an algorithm would take the continuous matrix product state on $N$ spins.  Such a state would be described by two operators: $H_L$ and $\hat Z$, where $\hat Z=S^z_N$ (we use the label $\hat Z$ here, rather than using $S^z_N$ as the
number of sites will change so we want to use a label that avoids specifying the site).  Then, we propagate the state one site to the right by adjoining an extra
site, giving a new system on $N+1$ sites.  We have the mapping:
\be
H_L \rightarrow H_L \otimes I + J \hat Z \otimes Z,
\ee
and
\be
\hat Z \rightarrow I \otimes Z,
\ee
where $I,Z$ denote the identity and $Z$ matrices on the $N+1$-st site.

For a practical algorithm, it is essential to truncate in some way.  We introduce a time dependent gauge transformation $X(t)$, replacing $\Lambda_b$ by $X(t) \Lambda_b X(t)^{-1}$.
Using this gauge transformation, the equation of motion becomes
\begin{eqnarray}
\partial_t \Lambda_b &=& - X(t) H_L X(t)^{-1} \Lambda_b - \Lambda_b (X(t)^{-1})^\dagger H_L X(t)^\dagger
\\ \nonumber
&& + |J| X(t) \hat Z X(t)^{-1} \Lambda_b 
 (X(t)^{-1})^\dagger \hat Z X(t)^\dagger
\\ \nonumber
&&+(\partial_t X(t)) X(t)^{-1} \Lambda_b + \Lambda_b (X(t)^{-1})^\dagger \partial_t X(t)^\dagger.
\end{eqnarray}
In general, the matrix $X(t)$ may be time dependent and may be non-unitary, so long as it is invertible.
The matrix $X$ is chosen to bring $\Lambda_b,\Lambda_t$ to diagonal matrices in the canonical form\cite{mpsreview}.  Once $\Lambda_b,\Lambda_t$ are both diagonal, the matrix $\Lambda$ is also diagonal, and we can then truncate to the subspace containing the largest eigenvalues.

In this particular problem, for long time the problem simplifies, as $\Lambda_b=\Lambda_t^\dagger$ and further $\Lambda_b$ becomes time independent.  As a result, it suffices to consider {\it unitary}, {\it time-independent} $X(t)$.  Since $X$ is unitary, the truncation then corresponds to truncating to a subspace
of the original Hilbert space.

We thus arrive at the following strange situation: for transverse evolution in imaginary time, in the long time limit, the algorithm is {\it equivalent} to the
original infinite-system DMRG algorithm, with one minor modification.  In the infinite system DMRG algorithm, we consider a system on $N$ sites.  We add an additional site to describe a new system on $N+1$ sites.  We then take two copies of that system on $N+1$ sites to get a system on $2N+2$ sites.  We compute the ground state density matrix, and use the largest eigenvalues of this density matrix to truncate the Hilbert space on the first $N+1$ sites.
Thus, this is the same as the transverse algorithm in this case, with the modification that the transverse algorithm considers a system on $2N+2$ sites with ferromagnetic coupling $|J|$, rather than considering the original Hamiltonian on $2N+2$ sites.  So, for example, the truncation would be different for the two algorithms if the Hamiltonian had an anti-ferromagnetic coupling.

Thus, there seems to be little point to consider the transverse algorithm here for practical applications of imaginary time simulations.  It is essentially the same as a known algorithm, except it uses the ``wrong" Hamiltonian in some circumstances.  However, in Refs.~\onlinecite{fold1,fold2,fold3} it was found that the transverse algorithm had advantages in other settings including real time evolution.  These other settings involve cases in which $\Lambda_b,\Lambda_t$ will be different, making the needed gauge transformations $X(t)$ non-unitary.  Later, we will develop a ``hybrid algorithm", which is based on the idea of combining the advantages of both approaches; namely, using the ``correct" Hamiltonian as in DMRG, but using more general gauge transformations and having different $\Lambda_b,\Lambda_t$.

This non-unitary choice of $X(t)$ gives more freedom in picking gauge transformations, increasing the power of the algorithm.
One way of understanding this power is to consider the toy model of Ref.~\onlinecite{fold3}.  In this model, entangled pairs are created and one particle in each pair propagates to the left at constant speed while the other propagates to the right at constant speed.  A direct calculation\cite{fold3} shows that the entanglement in the folded algorithm is bounded.  However, another way to understand it is to think about $\Lambda_b,\Lambda_t$.
In this case, $\Lambda_b$ is a high entropy state: all the left propagating degrees of freedom within distance $t$ of the cut are maximally mixed.
Similarly $\Lambda_t$ is a high entropy state, but in this case it is the {\it right}-propagating degrees of freedom near the cut that are maximally mixed.
This difference in which degrees of freedom are mixed allows a low bond dimension description of the transverse state.

There is however one interesting theoretical aspect of the transverse algorithm for imaginary time.  Suppose we have a finite size and we consider some cut across a vertical column of bond.  Let $\Psi_L$ and $\Psi_R$ be the left and right transverse states for that cut.  Thus, the desired partition function can be written as $\langle \overline \Psi_L | \Psi_R \rangle$,
where the inner product is in the space of bond variables and where we insert a complex conjugation symbol ($\overline{...}$) over $\Psi_L$ as the inner product by definition also complex conjugates the first variable.  Suppose we normalize so that $|\Psi_L|=|\Psi_R|=1$.  Then, suppose we truncate $\Psi_L$ to a lower bond dimension across some cut in time, approximating it by some state $\Psi_L^{trunc}$ with $|\Psi_L-\Psi_L^{trunc}|\leq \epsilon_L$ and similarly approximate $\Psi_R$ by $\Psi_R^{trunc}$ with $|\Psi_R-\Psi_R^{trunc}| \leq \epsilon_R$.  This leads to some error in the partition function.  However, by Cauchy-Schwarz we have
$\langle \overline \Psi_L^{trunc} | \Psi_R^{trunc} \rangle - \langle \Psi_L | \Psi_R \rangle \leq \epsilon_L + \epsilon_R + \epsilon_L \epsilon_R$.
Thus, since the tranverse algorithm will tend to pick truncations to minimize $\epsilon_L,\epsilon_R$, it will pick a truncation of the left state that will work to some accuracy for any choice of the right state and vice-versa.

\subsection{Evolution In Real Time}
We now consider how the entanglement grows with time for real-time evolution.  We focus on the case of unfolded evolution, and we consider the special case of
entanglement between the forward and return contour (i.e., between the top and bottom halves of the state).  By considering this case, we have $\Lambda_b=\Lambda_t^\dagger$, so it simplifies the problem.  As in the case of imaginary time evolution, we can obtain the correct result for the entanglement in time by treating $\Lambda_b$ as a pure state on a system of size $2N$ and computing the entanglement between left and right halves.

In Ref.~\onlinecite{fold3}, it was found that the entanglement grew linearly in time in the unfolded case for several examples where the system was started in its ground state at the bottom row of the tensor network.  To check this further, we considered the entanglement in a free fermion system as a model problem, equivalent to the Ising model
at vanishing parallel field.
Ref.~\onlinecite{fold3} considered the maximum entanglement (i.e., the entanglement across the cut that produces the largest entanglement), while
we consider just the case of entanglement between the forward and return contours.

We consider a translationally invariant free fermion problem: $H=\sum_i \psi^\dagger_i \psi_{i+1} + h.c.$, where $\psi^\dagger, \psi$ are
fermionic creation and annihilation operators.
We study the evolution of $\Lambda_b$ by replacing the coupling halfway across the system with an imaginary coupling: for a system of $2N$ sites, with $N$ sites in each half, we use a different Hamiltonian $\tilde H=\sum_{i<N}  \Bigl( \psi^\dagger_i \psi_{i+1} + h.c. \Bigr) - \sum_{i>N} \Bigl( \psi^\dagger_i \psi_{i+1} + h.c \Bigr)
+i (\psi^\dagger_N \psi_N + \psi_N \psi^\dagger_N)$ to describe the evolution of the state $\Phi$, $\partial_t \Phi=-i \tilde H \Phi$, where as above,
$\Phi$ is the pure state on $2N$ sites corresponding to $\Lambda_b$.

The problem of starting the system in its ground state at the bottom of the tensor network is not exactly the same problem as the problem of starting $\Phi$ in the ground state of either $H$ or $\tilde H$.  Instead, we need to construct a matrix product state describing the ground state of $H$, and then cut that state on the bond between the $N$-th and $N+1$-st site, and then combine that cut state with its adjoint on sites $N+1,...,2N$ to form a state on $2N$ sites.
The exact resulting state will depend upon the particular matrix product representation of the ground state of $H$ that we choose.
So, for simplicity, we choose instead to start the system in the ground state of the Hamiltonian 
$\sum_{i \neq N} \psi^\dagger_i \psi_{i+1} + h.c.$; that is the Hamiltonian corresponding to two decoupled systems of size $N$.
Similar results were found for a variety of other initial conditions.

What we find is that the entanglement entropy of $\Phi$ grows linearly in time.  However, if we consider evolution under either $H$ or under 
$\sum_{i<N}  \Bigl( \psi^\dagger_i \psi_{i+1} + h.c. \Bigr) - \sum_{i\geq N} \Bigl( \psi^\dagger_i \psi_{i+1} + h.c \Bigr)$, in both cases we find that
the entanglement grows only logarithmically in time.  The case of evolution under $H$ is the case of connecting two systems started in their ground state
by adding an additional coupling, and conformal field theory results indeed give this logarithmic coupling\cite{cft}.  The case of evolution under 
$\sum_{i<N}  \Bigl( \psi^\dagger_i \psi_{i+1} + h.c. \Bigr) - \sum_{i\geq N} \Bigl( \psi^\dagger_i \psi_{i+1} + h.c \Bigr)$ is slightly different as in this case
we connect two systems, one started in its ground state and the other started in its most excited state (we started in the ground state with a positive sign on the hopping term, but then evolved with a negative sign of the hopping term), but again we find numerically that the entanglement entropy grows only logarithmically with time.
Thus, we believe that the crucial reason for the linear entanglement growth is indeed the imaginary coupling.

\begin{figure}
\includegraphics[width=4in,angle=270]{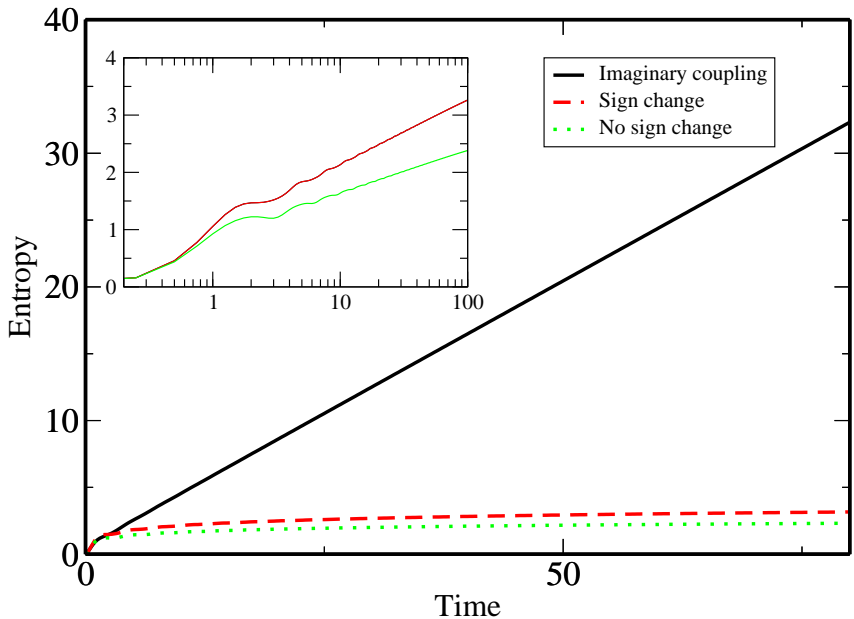}
\caption{Entanglement entropy as a function of time for free fermi system.  Top line is entropy using Hamiltonian $\tilde H$, while bottom two lines corresponding to using a Hermitian Hamiltonian (either the original Hamiltonian with couplings having changed sign to the right of the middle, or the original Hamiltonian).  Inset shows logarithmic scale for time axis, showing logarithmic growth of entropy for the latter two cases.}
\label{fffig}
\end{figure}

We briefly explain how we study the evolution of $\Phi$ under $\tilde H$.  Since $\tilde H$ is a non-Hermitian Hamiltonian, this is slightly more complicated than the usual free fermion problem.  The initial state is a Slater determinant with $N$ particles
and so can be written as
\be
\Phi(t=0)=
\Bigl( \prod_{a=1}^N (\sum_i A_{ia} \psi^\dagger_i)  \Bigr) |0\rangle,
\ee
where $|0\rangle$ is the vacuum state and $A$ is an $2N$-by-$N$ matrix.  We orthonormalize the rows of $A$ so that $A^\dagger A=I$.  Let $G$ be the Green's function matrix, whose elements are given by $G_{ij}=\langle \Phi(t=0) | \psi^\dagger_i \psi_j | \Phi(t=0) \rangle$.  Then, $G=A A^\dagger$.
After time $t$, the state evolves to change $A$ to $\exp(-i \tilde H t) A$.  However, since $\tilde H$ is non-Hermitian, the matrix $\exp(-i \tilde H t) A$ does not have orthonormal rows, and instead $G$ is equal to the projector onto the range of $A$.  So, to correctly compute the Green's function, we evolve $A$ and then orthonormalize in each time step, and then use this to compute $G$.  The entanglement entropy can be computed directly from the Green's function\cite{ffee}.

\section{Folding Hybrid Methods}
In this section we explain the hybrid algorithm and use it to study relaxation in an Ising model with combined transverse and parallel fields, comparing the original folding algorithm, the hybrid algorithm, and the iTEBD algorithm.
The idea of the hybrid algorithms is to maintain the advantage of the folding algorithm while removing the disadvantage of using the ``wrong Hamiltonian"
to evolve $\Lambda_b$.

\subsection{Hybrid Algorithms}
We begin by outlining our implementation of the standard folding algorithm before describing the hybrid algorithm.  We used a 
tensor network with discrete time, giving a second order Trotter-Suzuki decomposition of the time evolution.
The algorithm is initialized with a matrix product state of bond dimension $4$ exactly describing the amplitude for a column of bonds with one spin to the left.
Then, on each step of the algorithm, we adjoin an additional site, increasing the bond dimension.  Then, if the bond dimension becomes too large, we truncate to a lower bond dimension.
Almost all matrix product algorithms face the question of approximating a state with a given bond dimension by some state of lower bond dimension.
In many cases, this is done by trying to iteratively optimize the overlap\cite{iterativeopt}.  However, for our numerical work in this paper, for the standard
folding algorithm we use a simpler method.  

We bring the matrix product state to a canonical form.  In this form, we ensure that $\Lambda_t$ is equal to the identity matrix.  To perform this normalization, we use the method of iterated singular value decompositions explained in step 1 of theorem 1 of Ref.~\onlinecite{mpsreview}.  We bring $\Lambda_b$ to a diagonal matrix.  After bringing the entire matrix product state to a canonical form, we then truncate each bond to the subspace containing the largest eigenvalues of $\Lambda_b$ (note: we did not re-compute $\Lambda_b$ across a given bond after doing the truncation on the bonds below, though that would perhaps have increased the accuracy of the algorithm).

In this algorithm, $\Lambda_b$ is determined iteratively.  Let $b_f,b_r$ denote the state on a particular horizontal bond, where $b_f$ denotes the state on the forward contour and $b_r$ denotes the state on the return contour (note that we are describing a folded algorithm, so that both contours are combined into a single bond).
Given $\Lambda_b$ for a given bond, and given a set of matrices $A(b_f,b_r)$ we map according to
\be
\Lambda_b \rightarrow \sum_{b_f,b_r} A(b_f,b_r) \Lambda_b A(b_f,b_r)^\dagger.
\ee
Then, after computing $\Lambda_b$, we perform an eigendecomposition to find the unitary rotation that diagonalizes it; we then perform this rotation and
truncate.

The modification for the folding algorithm is simple.
We instead evolve according to
\be
\Lambda_b \rightarrow \sum_{b_f,b_r} A(b_f,b_r) \Lambda_b A(b_f,b_r)^T.
\ee
(This is done after bringing $\Lambda_t$ to the same canonical form as above).
The matrix $\Lambda_b$ is then not necessarily Hermitian.  Instead of performing an eigendecomposition, we perform a singular value
decomposition $\Lambda_b = U S V$.  We then rotate the basis so that $U$ is equal to the identity matrix and truncate to the subspace containing the largest
singular values.  This is equivalent to saying that we regard $\Lambda_b$ as a pure state on a system of $2N$ sites, and we truncate to the subspace
containing the largest eigenvalues of the density matrix on the first $N$ sites.

This essentially completes the description of the algorithm.  We run the transverse evolution until a fixed point is reached; one obvious improvement that we did not yet do is to do the transverse evolution at low bond dimension until a fixed point is reached and then increase the bond dimension, again iterate to a fixed point, and so on.  Another obvious question is what accuracy is obtained if the tensor network for a given total time $t$ is used to evaluate observables at some time $t'<t$; that is, if the observable is placed not at the middle of the tensor network but somewhere in either the forward or reverse contour.
Using this, one could with almost the same numerical effort as is required to compute the expectation value at time $t$ also compute the expectation value for all times $t'<t$ as the only additional numerical effort required would be at the last stage when the expectation value is evaluated, with the computation of the transverse evolution being the same.  We did not do this since our focus was on testing the accuracy of the original and hybrid methods, and trying (later) to compute accurate results at a few particular times for certain initial conditions to study long time behavior and support predictions from iTEBD.

The reader will notice that this procedure is close to the original infinite system version of DMRG.  The system grows from one step to the next, so that, for example, if we have $N=3$, we find the largest eigenvalues of the density matrix of a system on $3+3=6$ sites.

\subsection{Hamiltonian and Numerical Methods}
The Hamiltonian we considered is the same as in Eq.~(\ref{Ham1}), studied in Ref.~\onlinecite{fold2}, namely an Ising model with transverse plus parallel fields,
with $J=1, H=-1.05, g=0.5$.  We considered two different initial conditions, either starting all spins in the $|X+\rangle$ state or starting
all spins in the $|X-\rangle$ state.  The latter choice is equivalent up to a unitary transformation (conjugating by $\prod_i S^z_i$) to starting in the $|X+\rangle$ state and picking $H=+1.05$.
We fixed the timestep $\dt=0.1$.

For the iTEBD algorithm, to obtain an accurate comparison of results for observables, it is necessary to employ the same Trotter-Suzuki decomposition
$H=A+B$, as before.
The iTEBD algorithm of Ref.~\onlinecite{itebd} employs a decomposition of the unitary operator $\exp(-i H \dt) \approx U_A U_B$ where now $U_A$ acts on pairs of sites connected by an even bond (i.e., it is a product of unitaries, each on a pair of sites $2k,2k+1$ for integer $k$) while $U_B$ acts on pairs of sites connected by an odd
bond (i.e., it is a product of unitaries, each on a pair $2k-1,2k$).  By choosing $U_A=\exp(-i A \dt/2) \exp(-i \sum_{k} J S^z_{2k} S^z_{2k+1} \dt)$ and
$U_B=\exp(-i \sum_k S^z_{2k-1} S^z_{2k} \dt) \exp(-i A \dt/2)$, the iTEBD algorithm replicates the Trotter-Suzuki decomposition used here
(note that the iTEBD algorithm describes the state as translationally invariant with period $2$, while in fact in this case exact evolution leaves the
state invariant with period $1$).

However, we observed that for this Trotter-Suzuki decomposition, the iTEBD algorithm as described in Ref.~\onlinecite{itebd} exhibited poor numerical stability associated with division by small singular values, even for very modest values of the bond dimension.  We do not have any insight into why this problem occurred at such small values of the bond dimension for this system when for most other systems it does not occur until the bond dimension is quite large.  However, we were able to
significantly ameliorate this problem by using the modification of the algorithm given in section IIA of Ref.~\onlinecite{lightcone} (see also section 7.3.2 of Ref.~\onlinecite{dmrgmps}).  If arithmetic were done exactly, the modified algorithm would be equivalent to the original iTEBD algorithm.  However, for finite precision arithmetic it is more stable as completely avoids division which can lead to instability when singular values can become small, at a slight additional computational cost of one extra matrix multiplication (this cost is small compared to the cost of singular value decompositions).

\subsection{Accuracy Results}
We now present a comparison of the hybrid folding method to the original folding method.  In addition to these two numerical techniques, we used an iTEBD
algorithm for comparison; at short times, the iTEBD is able to provide very accurate values of certain observables, while at long times the folding methods
are more accurate and we instead compare the results at small bond dimension to the results at large bond dimension; additional error estimates are obtained by measuring the error in the identity observable: we know that the expectation value of the identity operator should be $1$, while for finite bond dimension the value may be slightly different.

We only consider errors for the case of the transverse plus parallel field Ising model (i.e., the case that $g\neq 0$) as in the case of $g=0$ both algorithms are extremely accurate.
A very heuristic explanation of this is as follows: for an infinite system, an excitation created at a given time may propagate to the left indefinitely, giving a behavior similar to that of the toy model of Ref.~\onlinecite{fold3} where the entanglement entropy in the time directions saturates at some finite value, even for large time.  Alternatively, the excitation can scatter off other excitations and ``bounce back", showing behavior different from the toy model, leading to an entanglement entropy growing with time (we discuss the entanglement entropy later).  For $g=0$, since the model maps to non-interacting fermions, perhaps it is closer to the toy model where this scattering off other excitations is absent.

We measured three different kinds of errors.  First, the error in the expectation value of the $X$ operator; we measured the $X$ operator by inserting
the $X$ operator at the middle of the tensor network on a single site, using the folded algorithm to compute that expectation value, and then normalized by dividing by the expectation value of the identity operator.
To do this, we propagated the matrix product state from the left (the propagation from the right is identical) until a fixed point was reached, and then sandwiched a single column containing either the $X$ operator or the identity operator between these states from the left and right, and then evaluated the resulting network (we remark that for longer times and smaller bond dimensions, we did not truly reach a fixed point as we observed some slight fluctuations in the observable within a narrow range; however, these fluctuations were smaller than the errors reported below).
We then compared this expectation value to iTEBD results at short times where iTEBD is accurate.  
We then repeated this error calculation for the $Z$ operator; results for that are not shown, but are similar.
Finally, we measured the error in the expectation value of the identity operator itself.  Of course, the expectation value of the identity operator
should be $1$, but truncation error causes it to differ from $1$.  However, the difference relative to $1$ increases with the number of columns, as each column will incur some truncation error.
We overcome this by computing the ratio of the expectation value of the identity operator for $N$ columns and $N+1$ columns, taking $N$ large enough to reach a fixed point.

All the results are plotted for an infinite system; more precisely, we evolve for a finite distance until we reach a fixed point with sufficiently high accuracy.
We do not show results for finite sizes, but anecdotally we observe an interesting phenomenon.  If we propagated for some number, $N$, of columns, and then compute the expectation value, this gives the value of $2N+1$ sites.  We found that in many cases the error was {\it larger} for small $N$ than for larger $N$.  Heuristically this may arise for the reason that for finite sizes an excitation propagating to the left can ``bounce off the edge", even in the non-interacting limit, leading to increased entanglement in the time direction (in fact, we observe this in the entanglement entropy, so the entanglement in time first increased with $N$, and then later decreased at larger $N$).  Another reason for error may be the fact that we are using an infinite system algorithm.

We consider both initializing in the $|X+\rangle$ state as in Ref.~\onlinecite{fold2} and initializing in the $|X-\rangle$ state (see next section for interesting physical behavior in this state).
In all cases, the hybrid algorithm is more accurate.
In several cases, the hybrid algorithm at given bond dimension is more accurate than the original folded algorithm at double the bond dimension.

\begin{figure}
\includegraphics[width=4in,angle=270]{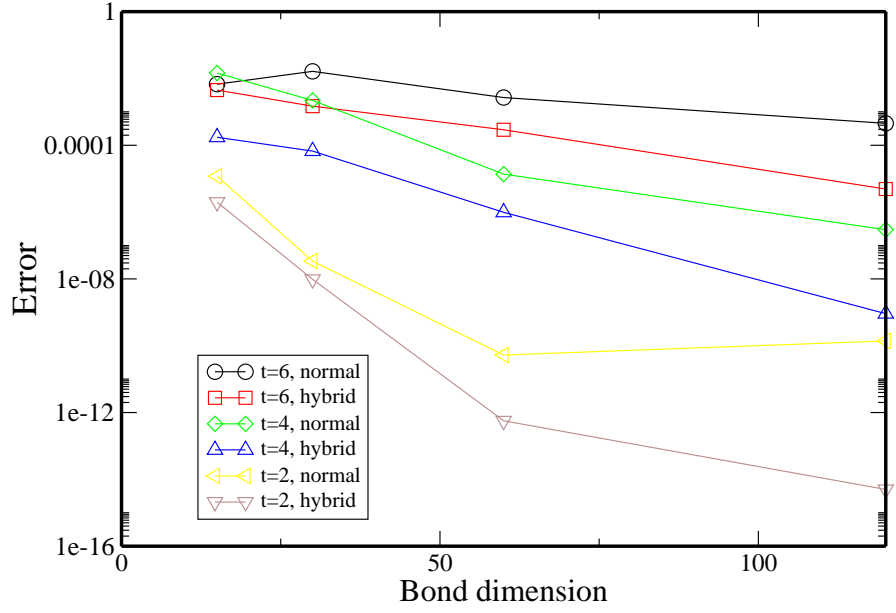}
\caption{Error in expectation value of $X$ operator, for times $t=2,4,6$ and for normal and hybrid algorithms.  Initial conditions are in $|X+\rangle$ state.}
\label{mXefig}
\end{figure}

\begin{figure}
\includegraphics[width=4in,angle=270]{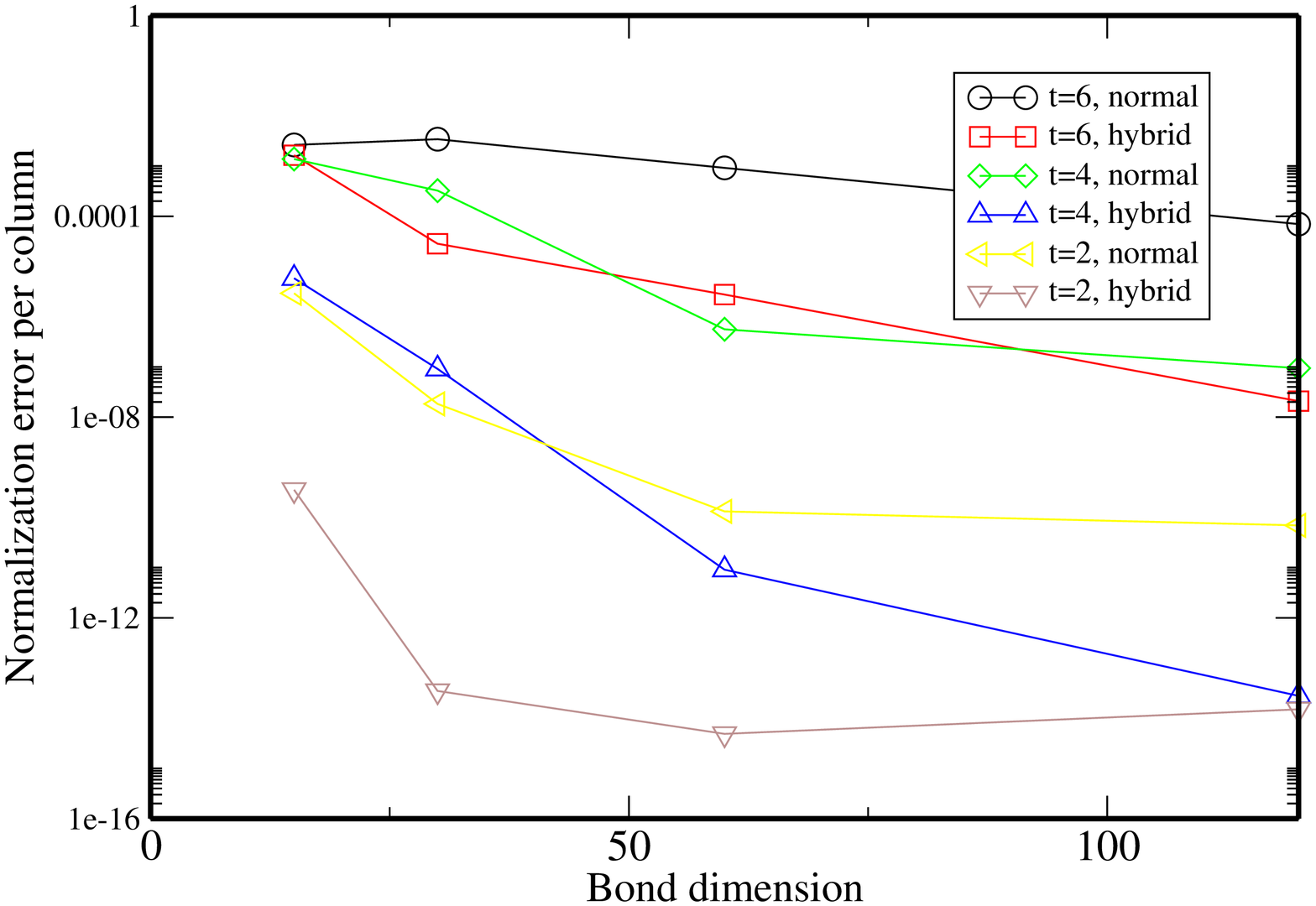}
\caption{Error (per column) in expectation value of identity, for times $t=2,4,6$ and for normal and hybrid algorithms. Initial conditions are in $|X+\rangle$ state.}
\label{mIefig}
\end{figure}

\begin{figure}
\includegraphics[width=4in,angle=270]{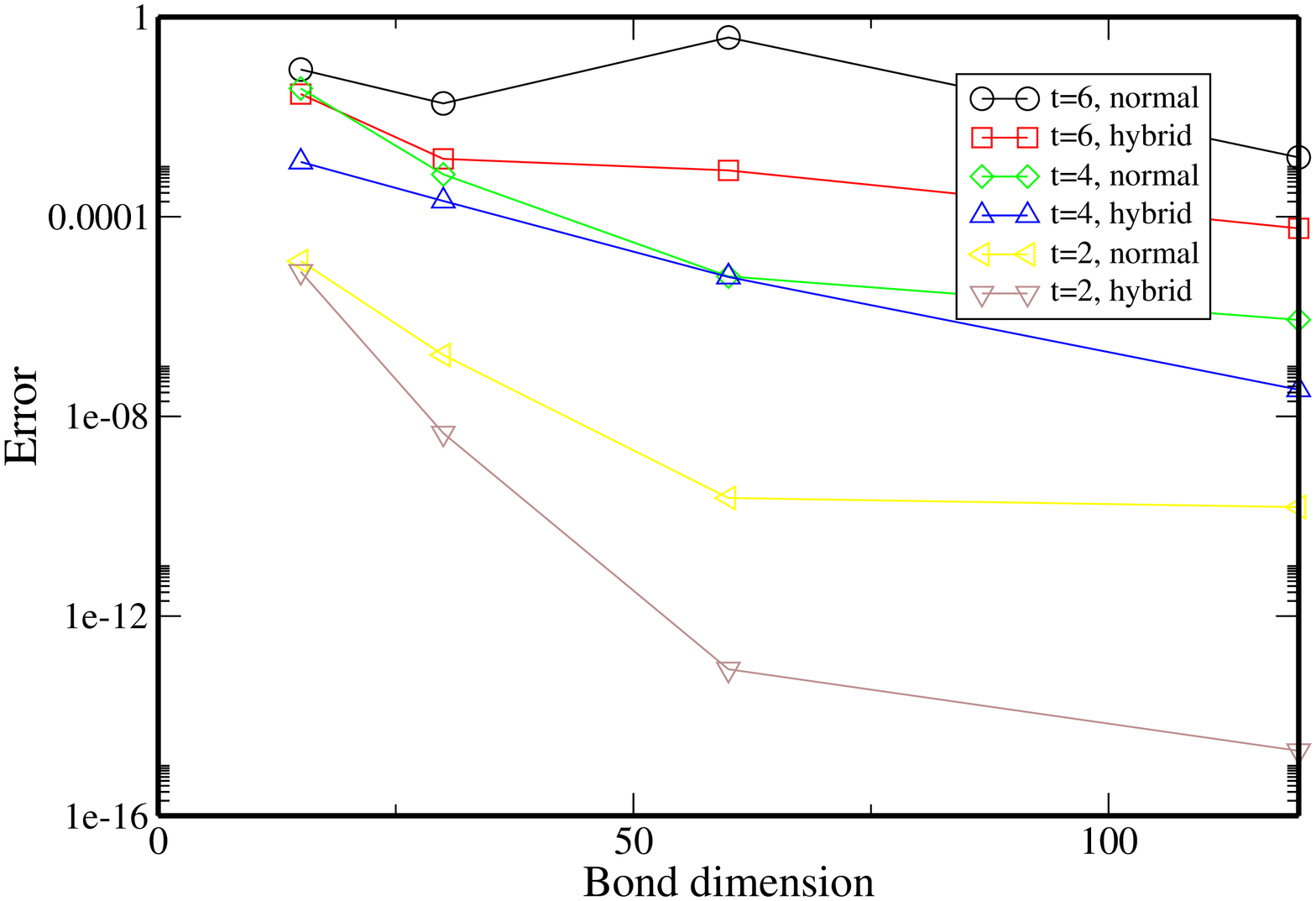}
\caption{Error in expectation value of $X$ operator, for times $t=2,4,6$ and for normal and hybrid algorithms.  Initial conditions are in $|X-\rangle$ state.}
\label{pXefig}
\end{figure}

\begin{figure}
\includegraphics[width=4in,angle=270]{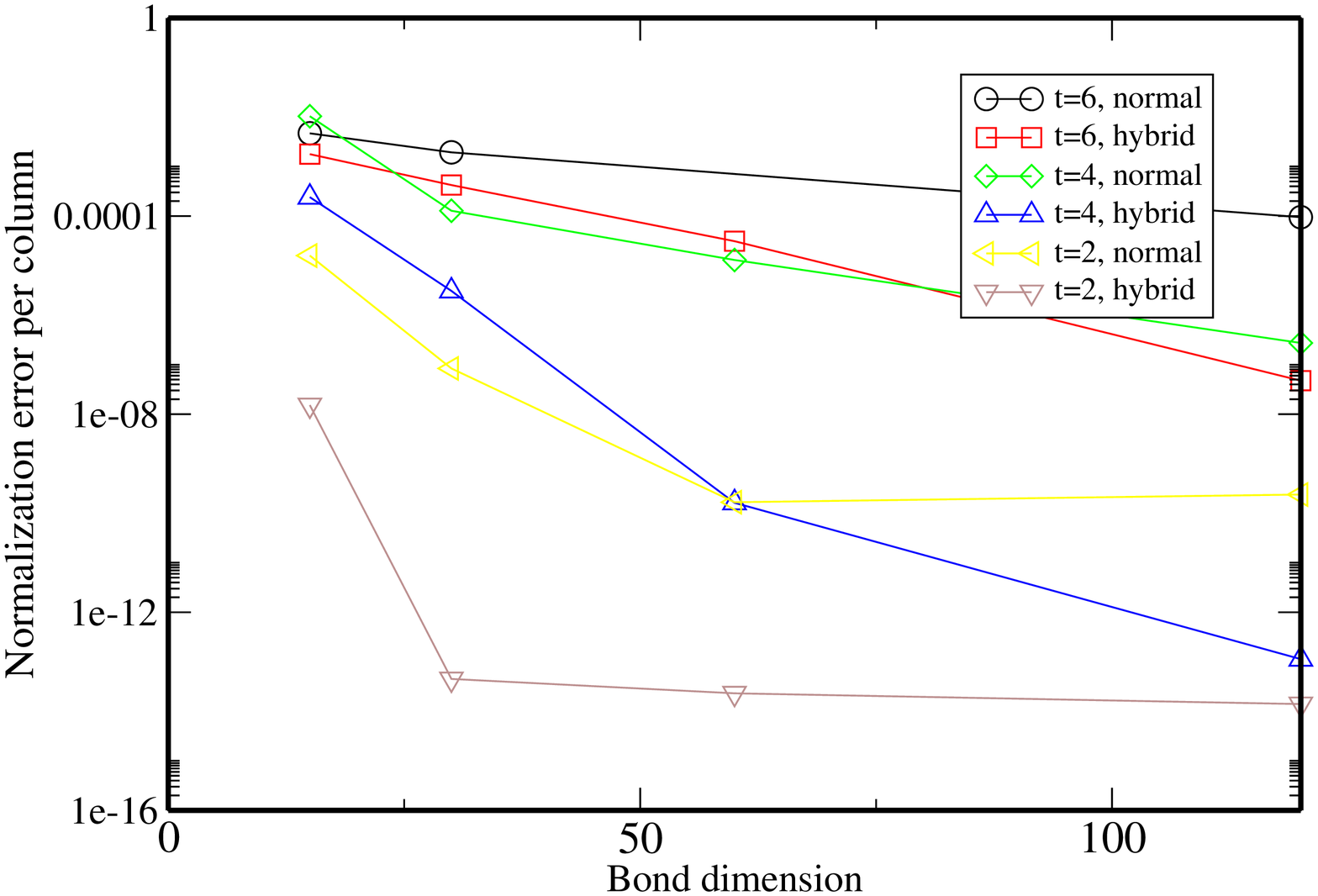}
\caption{Error (per column) in expectation value of identity, for times $t=2,4,6$ and for normal and hybrid algorithms. Initial conditions are in $|X-\rangle$ state.}
\label{pIefig}
\end{figure}

Figures ~\ref{mXefig},\ref{mIefig} show the errors for the $|X+\rangle$ initial state, while
Figures ~\ref{pXefig},\ref{pIefig} show the errors for the $|X-\rangle$ initial state.

\section{Long Time Oscillations in $|X-\rangle$ State}
\begin{figure}
\includegraphics[width=4in,angle=270]{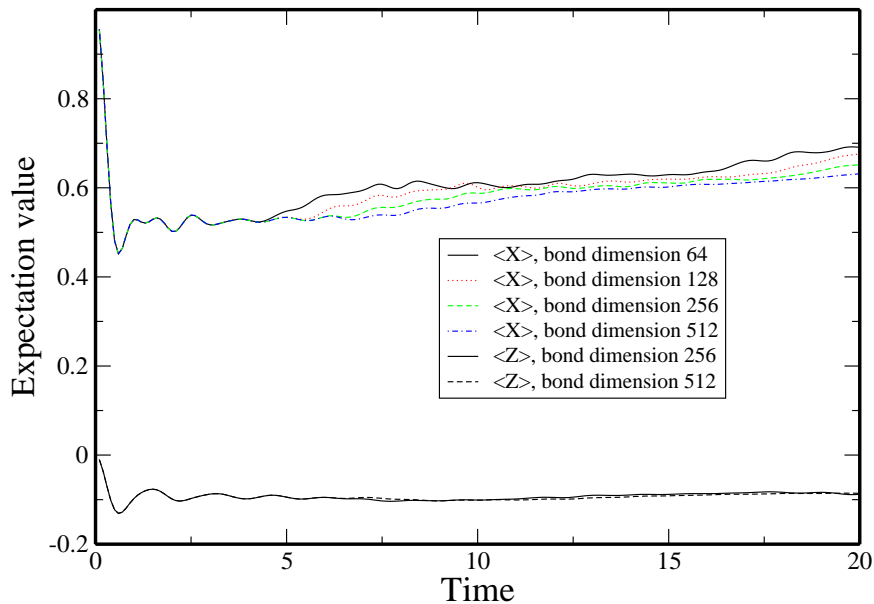}
\caption{Expectation value of $X$ and $Z$ from iTEBD for various bond dimensions. Initial conditions are in $|X+\rangle$ state.}
\label{itminus}
\end{figure}

We now consider the long time behavior.  We study both $|X-\rangle$ and $|X+\rangle$ states, but the most interesting physics arises in the $|X-\rangle$
initial state.
Figure ~\ref{itminus} shows the behavior of $\langle X \rangle$ and $\langle Z \rangle$ from the iTEBD algorithm for a variety of bond dimensions for
the $|X+\rangle$ initial state.
As seen in Ref.~\onlinecite{fold2}, the iTEBD results become unreliable for long time.  In particular, the expectation value $\langle X \rangle$ becomes very inaccurate at long time.  The folding algorithm is able to provide reliable results at times that the iTEBD algorithm cannot reach.  In fact, this kind of behavior occurs for
many systems subject to a global quench: at a given bond dimension the iTEBD algorithm is extremely quantitatively accurate until a certain time, after which the results become {\it qualitatively} incorrect.  The results from Ref.~\onlinecite{fold2} indicate that the actual values $\langle X \rangle$ and $\langle Z \rangle$ converge in the long time limit, becoming approximately stationary as a function of time.

\begin{figure}
\includegraphics[width=4in,angle=270]{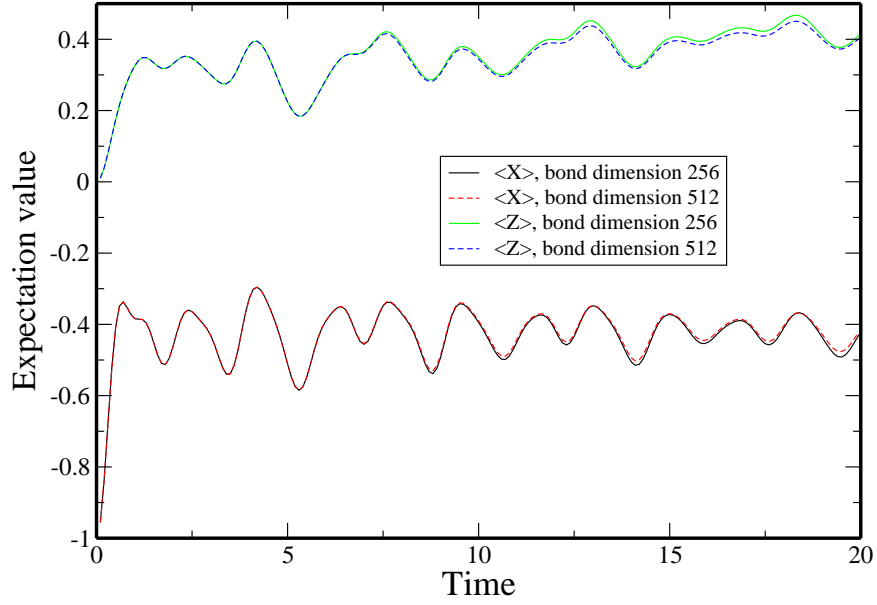}
\caption{Expectation value of $X$ and $Z$ from iTEBD for various bond dimensions. Initial conditions are in $|X-\rangle$ state.}
\label{itplus}
\end{figure}

Figure ~\ref{itplus} shows the situation for the $|X-\rangle$ initial state.  Here the situation is almost exactly the opposite.  First, even when the time is long enough that the iTEBD results are quantitatively incorrect, they continue to agree qualitatively.  We only show bond dimensions $256$ and $512$, but others are similar.  Second, if the iTEBD results are to be trusted, rather than $\langle X \rangle$ and $\langle Z\rangle$ converging to some value, they instead show quasi-periodic long time oscillations.  Note that both the $256$ and $512$ simulations show almost identical locations for the peaks of $\langle X \rangle$ and $\langle Z \rangle$.

\begin{figure}
\includegraphics[width=4in,angle=270]{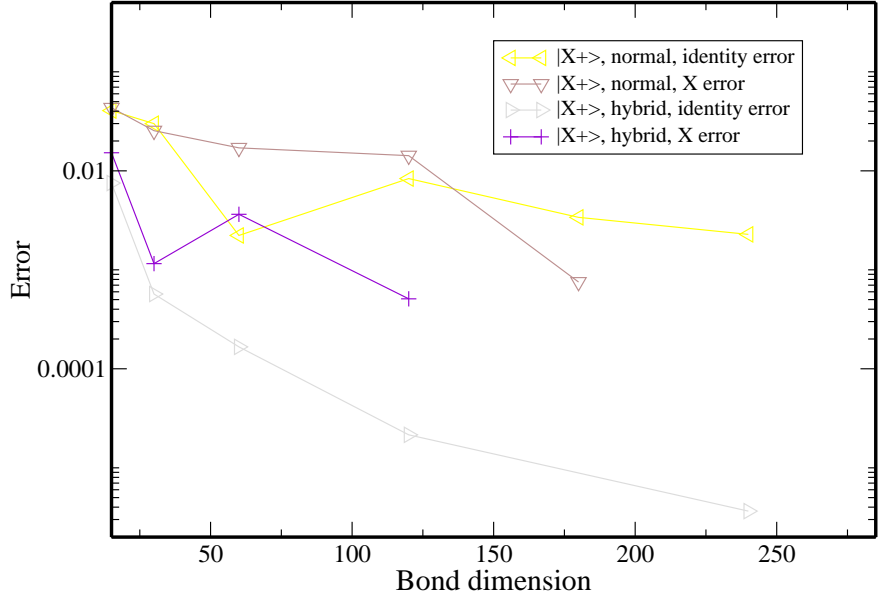}
\caption{Errors in measurement of $\langle X \rangle$ and for expectation value of identity operator, for both methods for $t=10.6$.  Errors for $\langle X \rangle$ are obtained by comparing to bond dimension $240$ runs.  Initial conditions are in $|X+\rangle$ state.}
\label{minus10}
\end{figure}

\begin{figure}
\includegraphics[width=4in,angle=270]{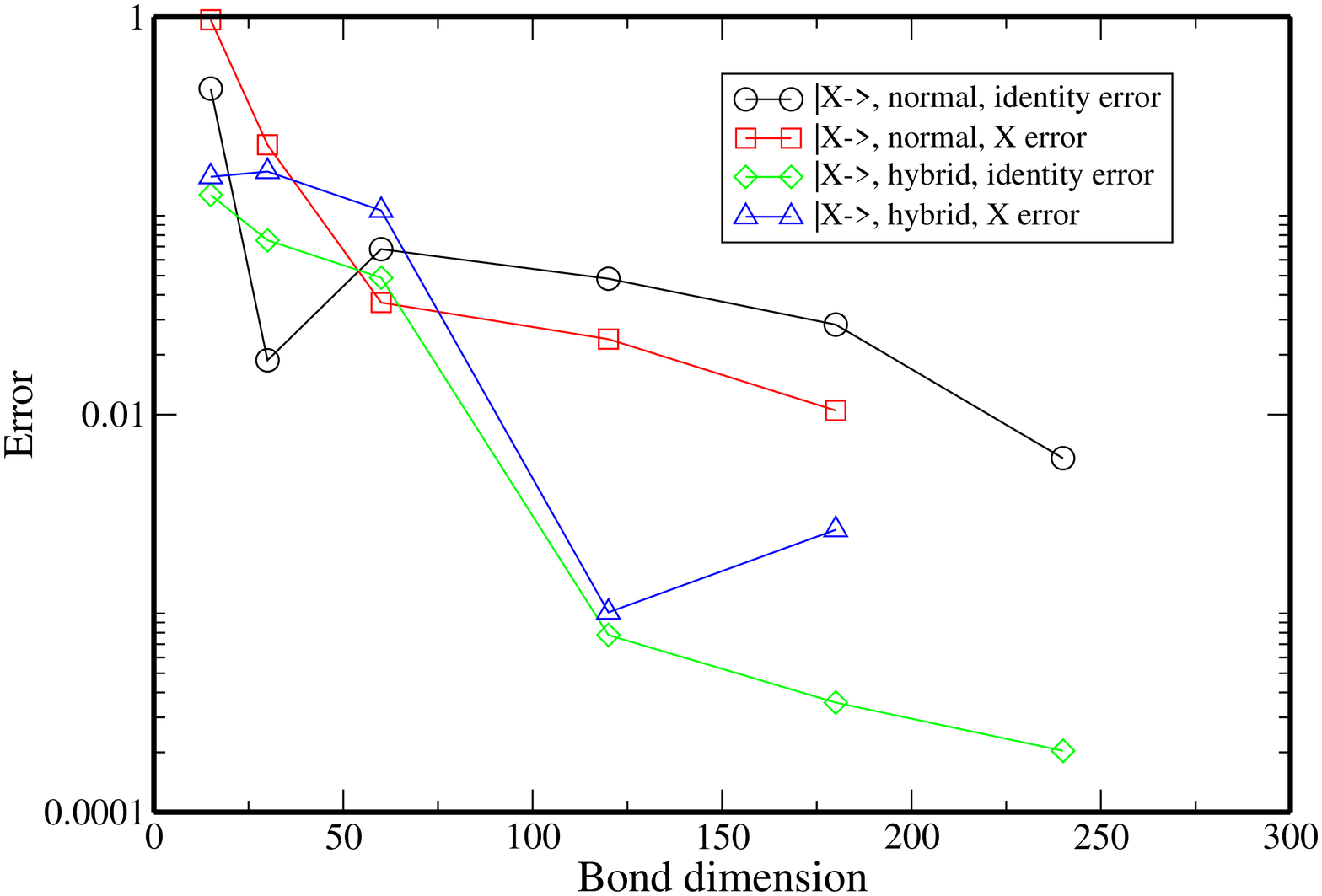}
\caption{Errors in measurement of $\langle X \rangle$ and for expectation value of identity operator, for both methods for $t=10.6$.  Errors for $\langle X \rangle$ are obtained by comparing to bond dimension $240$ runs.  Initial conditions are in $|X-\rangle$ state.}
\label{plus10}
\end{figure}

In this section, we probe some of this long time behavior with the folding algorithm and find that it supports the possibility that these long time oscillations are real.  We focus on two specifics times, $t=10.6$ and $t=14.0$, close to peaks of the expectation value of $X$ in Fig.~\ref{itplus}.  We continue to find that the hybrid method is more accurate than the normal method.  We do not have sufficiently accurate iTEBD data to compare to; it is likely possible to obtain this data using present resources at time $10.6$ for the $|X-\rangle$ initial state but likely not possible for the $|X+\rangle$ initial state or for either state at time $t=14.0$.  Hence, to measure the error in the expectation value of $X$ for a given method (hybrid or normal) and given initial conditions, we compare the difference between the value obtained for a given bond dimension for $\langle X \rangle$ and the value obtained for bond dimension $240$ for $\langle X \rangle$.
Figs.~\ref{minus10},\ref{plus10} show that respective errors.  Note that the error in $\langle X \rangle$ is usually larger than the error (per column) in the expectation value of the identity operator, but is comparable.

Based on these results, the hybrid method simulations are likely accurate to with an error $\sim 2\times 10^{-3}$ at time $10.6$ for the $|X-\rangle$ intial conditions.  The value of $\langle X \rangle$ obtained is $0.4785...$ at bond dimension $240$.  This is to be compared to the iTEBD result at bond dimension $512$ of $0.4890...$.  We believe that the hybrid method is more accurate; however, it supports the qualitative conclusion that the oscillations observed in the iTEBD data continue for these long times as the hybrid result gives a value of $\langle X \rangle$ which is significantly above the mean value at long times.

We now turn to results at time $t=14.0$.  In this case, we only ran the hybrid method and we only ran the $|X-\rangle$ initial conditions.  At bond dimension $180$ the identity error was $0.0012...$ while at bond dimension $240$ it was $0.0014...$.  The difference in the expectation value $\langle X \rangle$ between bond dimension $180$ and $240$ was $0.0046...$, giving some indication of the accuracy of results in this case.  The value of $\langle X \rangle$ was $0.472...$ for bond dimension $240$.  This is to be compared to $0.494...$ from iTEBD at bond dimension $512$.  In this case, again some quantitative difference is observed between the hybrid and iTEBD results, but the hybrid results support the notion that long times oscillations persist for the $|X-\rangle$ initial state.

\begin{figure}
\includegraphics[width=4in,angle=270]{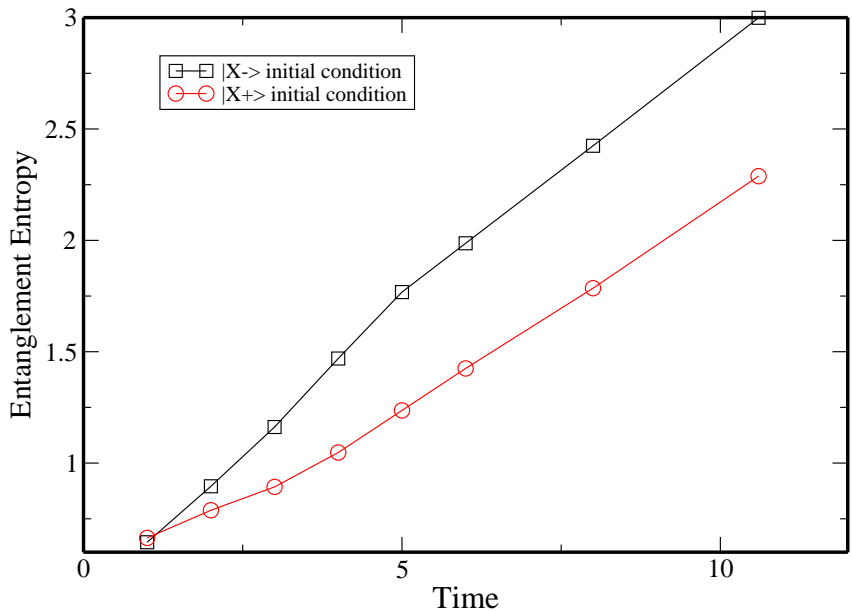}
\caption{Entanglement entropy in time for $|X-\rangle$ and $|X+\rangle$ initial conditions as a function of time.  Results are from bond dimension $120$ simulations with the original folding algorithm.}
\label{entanglementgraphfig}
\end{figure}

We emphasize that we do not understand the reason for these oscillations and only have numerical data to support their existence.
However, as a final comparison of the difference between the $|X+\rangle$ and $|X-\rangle$ initial conditions, we plot in Fig.~\ref{entanglementgraphfig} the entanglement of the matrix product state of the folding evolution, as a function of time.  The entropy plotted is the maximum, over all cuts of the wavefunction.
The simulations are from the original folding algorithm with bond dimension $120$ (somewhat surprisingly, the entropy actually {\it decreased} with increasing bond dimension, and it is not fully stabilized at the longest times for bond dimension $120$ as the results for dimension $240$ differ in the second decimal place; thus the results are not fully quantitatively accurate).
Both initial conditions show a linear growth with time (contrasted with several cases in Ref.~\onlinecite{fold3} where the entropy saturated as a function of time), but for the $|X-\rangle$ initial condition the slope is distinctly greater.  This likely explains why the simulations for $|X-\rangle$ are more difficult.  It would be interesting to probe the evolution of the entanglement entropy for a variety of initial conditions and see when it is linear in time and when not.

\section{Discussion}
We have studied the temporal entanglement in the folding algorithm, and shown that it can be understood as spatial entanglement in a modified
Hamiltonian.  This suggested the modification to the hybrid algorithm above.  Using this algorithm, we have studied longer time dynamics in a non-integrable
spin chain and provided additional numerical evidence for persistent oscillations under certain initial conditions.  The physical origin of these oscillations remains unclear to us.

As remarked, the hybrid algorithm that we have implemented bears some similarity to the growing infinite system DMRG.  An obvious question is whether
the algorithm can work for finite, translationally noninvariant systems.  A natural way to do this for a system of $M$ total sites would be to store, for each $N$,
matrix product states $\Psi_N^{L},\Psi_N^R$ describing the left and right transverse states after cutting across the $N$-th column.  Then, one could sweep right, generating a new $\Psi_{N+1}^L$ from $\Psi_N^L$ and use $\Psi_{N+1}^R$
to guide the truncation of $\Psi_{N+1}^L$, followed by a sweep left, generating a new $\Psi_{N-1}^R$ from $\Psi_N^R$ and use $\Psi_{N-1}^L$ to guide
the truncation of $\Psi_{N-1}^R$.
To guide this truncation, one could employ a natural analogue of the hybrid algorithm.  To truncate $\Psi_{N+1}^L$, one could first compute a $\Lambda_t$ from $\Psi_{N+1}^L$ and make a gauge transformation to bring $\Lambda_t$ to the identity matrix; a similar gauge transformation could be applied to $\Psi_{N+1}^R$ to bring $\Lambda_t$ for that state also to the identity matrix.  One could then evolve $\Lambda_b$ according to
\be
\Lambda_b \rightarrow \sum_{b_f,b_r} A^L(b_f,b_r) \Lambda_b \Bigl(A^R(b_f,b_r)\Bigr)^T,
\ee
where $A^L,A^R$ are the matrix in the matrix product states $\Psi_{N+1}^L,\Psi_{N+1}^R$.

We leave a study of this algorithm for future work.  One might also wonder whether we can more accurately treat $\Lambda_t$.  That is, currently we evolve $\Lambda_b$ according to the ``correct Hamiltonian", while $\Lambda_t$ is modified according to a modified Hamiltonian.  This is also a task for the future.

Finally, we briefly speculate about whether an algorithm that works directly with the continuous matrix product state could be practical.  
The continuous matrix product state for an unfolded system is parameterized by two (possibly time-dependent) matrices, $H_L(t), \hat Z(t)$.
In the folded case, we need three matrices, $H_L(t), \hat Z_f(t), \hat Z_r(t)$.
It seems impractical to apply arbitrary time dependent gauge transformations $X(t)$.  So, one option would be to apply piecewise-constant gauge transformations, breaking the total time evolution into discrete time intervals and keeping $X(t)$ constant on each interval.  Then it will not be possible to make $\Lambda_b(t),\Lambda_t(t)$ both diagonal for all times $t$, however we can make them both diagonal at, for example, the middle of each time interval, and use this to perform a truncation.  The last step needed to complete this continuous algorithm is a way to solve equations of motion similar to
Eq.~(\ref{timeev}), with $S^z_N$ replaced with the appropriate operators $\hat Z$ (or, in the folded case, $\hat Z_f,\hat Z_r$).
Such a linear equation of motion can be integrated to very high accuracy over short time intervals; for example, a series expansion for the solution can
readily make the error in evolving this equation of motion exponentially small in the order.  Hence, in this way we can make the discretization
error present in Trotter-Suzuki methods exponentially small, even for relatively large time length for each interval.  This method would seem to have two main disadvantages (beyond of course the complexity of the method).  First, while mathematically
it is convenient to describe choosing a gauge transformation to make $\Lambda_b,\Lambda_t$ diagonal, in practice we would like to be able to perform this
transformation in a numerically stable way.  The most naive way of doing this (doing an eigendecomposition on $\Lambda_t$, rotating so that $\Lambda_t$ is diagonal, and then performing another gauge transformation by a diagonal matrix containing the singular values of $\Lambda_t$, finally followed by a rotation to make $\Lambda_b$ diagonal)) is not numerically stable due to to division by small singular values; we would prefer a method like that in Ref.~\onlinecite{mpsreview} as used here which uses iterated singular value decompositions and avoids divisions.  However, we do not currently know
how to perform the gauge transformation in such a stable way for the continuous state.  The second disadvantage is that we can only pick the gauge transformation to make $\Lambda_b,\Lambda_t$ diagonal at certain discrete times, thus possibly increasing the truncation error; this disadvantage would be reduced if we took the time of each interval short (as the $\Lambda$ matrices would then change only a small amount over an interval), but if we are going to take short intervals then a discrete method will have small Trotter-Suzuki error and so there would be little advantage.


\begin{thebibliography}{99}
\bibitem{dmrg} S. R. White, Phys. Rev. Lett. {\bf 69}, 2863 (1992),
S. R. White, Phys. Rev. B {\bf 48}, 10345 (1993).

\bibitem{tebd} G. Vidal, Phys. Rev. Lett. {\bf 91}, 147902 (2003); G. Vidal, Phys. Rev. Lett. {\bf 93}, 040502 (2004).

\bibitem{tdmrg} A. J. Daley, C. Kollath, U. Schollw\"{o}ck, and G. Vidal, J. Stat. Mech., P04005 (2004);
S. R. White and A. E. Feiguin, Phys. Rev. Lett {\bf 93}, 076401 (2004).


\bibitem{therm3} J. Burges and J. Cox, ``Thermalization of quantum fields from time-reversal invariant evolution equations", Phys.
Rev. A {\bf 517}, 369 (2001).

\bibitem{therm2} J. M. Deutsch, ``Quantum statistical mechanics in a closed system",
Phys. Rev. A {\bf 43}, 2046 (1991).

\bibitem{therm1} P. Calabrese and J. Cardy, ``Time-dependence of correlation functions following a quantum quench",
 Phys.Rev.Lett. {\bf 96}, 136801 (2006).

\bibitem{therma} M. Srednicki, Phys. Rev. E {\bf 50}, 888 (1994)

\bibitem{therm0} M. Rigol, V. Dunjko, and M. Olshanii, ``Thermalization and its mechanism for generic isolated quantum systems",
Nature {\bf 452}, 854 (2008).

\bibitem{thermf} M. Cramer, C.M. Dawson, J. Eisert, T.J. Osborne, ``Exact relaxation in a class of non-equilibrium quantum lattice systems",
Phys. Rev. Lett. {\bf 100}, 030602 (2008).

\bibitem{dmrgmps} U. Schollw\"{o}ck, ``The density-matrix renormalization group in the age of matrix product states",
Annals of Physics {\bf 326}, 96 (2011).

\bibitem{entropy} P. Calabrese and J. Cardy, ``Entanglement Entropy and Quantum Field Theory",
J.Stat.Mech.0406:P06002,2004.

\bibitem{eb1} A. M. Childs, D. W. Leung, F. Verstraete, and G. Vidal,
Quantum Information and Computation {\bf 3}, 97 (2003).

\bibitem{entropybound} S. Bravyi, M. B. Hastings, F. Verstraete, 
``Lieb-Robinson bounds and the generation of correlations and topological quantum order",
Phys. Rev. Lett. {\bf 97}, 050401 (2006).

\bibitem{eb2} S. Bravyi, ``Upper bounds on entangling rates of bipartite Hamiltonians", Phys. Rev.
A, {\bf 76}, 052319, Nov 2007.

\bibitem{eb3} M. Mari\"{e}n, K. M.R. Audenaert, K. Van Acoleyen, and F. Verstraete,``Entanglement Rates and the Stability of the Area Law for the Entanglement Entropy", arXiv:1411.0680.

\bibitem{fold1} M. C. Ba\~{n}uls, M. B. Hastings, F. Verstraete, and J. I. Cirac, ``Matrix Product States for dynamical simulation of infinite chains",
Phys. Rev. Lett. {\bf 102}, 240603 (2009).

\bibitem{fold2} M. C. Ba\~{n}uls, J. I. Cirac, and M. B. Hastings, ``Strong and weak thermalization of infinite non-integrable quantum systems",
Phys. Rev. Lett. {\bf 106}, 050405 (2011).

\bibitem{fold3} Alexander M\"{u}ller-Hermes, J. Ignacio Cirac, Mari Carmen Ba\~{n}uls, ``Tensor network techniques for the computation of dynamical observables in 1D quantum spin systems", New J. Phys. {\bf 14} 075003 (2012).

\bibitem{itebd} G. Vidal, ``Classical simulation of infinite-size quantum lattice systems in one spatial dimension",
Phys. Rev. Lett. {\bf 98}, 070201 (2007).

\bibitem{lightcone} M. B. Hastings, ``Light Cone Matrix Product", J. Math. Phys. {\bf 50}, 095207 (2009).


\bibitem{mpsreview} D. Perez-Garcia, F. Verstraete, M.M. Wolf, and J.I. Cirac, ``Matrix Product State Representations",
Quantum Inf. Comput. {\bf 7}, 401 (2007).

\bibitem{CMPS} F. Verstraete and J. I. Cirac, ``Continuous matrix product states for quantum fields", Phys. Rev. Lett. {\bf 104}, 190405 (2010).

\bibitem{cft} P. Calabrese and J. Cardy, ``Entanglement and correlation functions following a local quench: a conformal field theory approach",
J. Stat. Mech., P10004 (2007).

\bibitem{ffee} J. Eisert, M. Cramer, M.B. Plenio, ``Area laws for the entanglement entropy - a review",
Rev. Mod. Phys. {\bf 82}, 277 (2010).

\bibitem{mporep} V. Murg, J.I. Cirac, B. Pirvu, F. Verstraete, ``Matrix product operator representations",
New J. Phys. {\bf 12}, 025012 (2010).

\bibitem{iterativeopt} See section 4.5.2 of Ref.~\onlinecite{dmrgmps}.
\end{thebibliography}
\end{document}